\pdfoutput=1
\documentclass[12pt]{article}
\usepackage{graphicx,amssymb,amsmath}
\usepackage{epsfig}
\usepackage{color,graphicx}
\usepackage{cite}
\usepackage{amssymb,amsmath}
\usepackage{multirow}
\usepackage{hyperref}
\usepackage[utf8]{inputenc}
\usepackage{subcaption}
\usepackage{tabularx}
\newcolumntype{Y}{>{\centering\arraybackslash}X}
\usepackage[top=2.5cm, bottom=2.5cm, left=2.5cm, right=2.5cm]{geometry}	

\numberwithin{equation}{section}

\newcommand{\be}{\begin{equation}}
\newcommand{\ee}{\end{equation}}
\newcommand{\bea}{\begin{eqnarray}}
\newcommand{\eea}{\end{eqnarray}}

\newcommand{\eps}{\epsilon}

\begin{document}
 
\begin{flushright}
HIP-2021-40/TH\\
APCTP Pre2021 - 033
\end{flushright}

\begin{center}

\centering{\Large {\bf Holographic QCD in the \emph{NICER} era}}

\vspace{5mm}

\renewcommand\thefootnote{\mbox{$\fnsymbol{footnote}$}}
Niko Jokela,${}^{1,2}$\footnote{niko.jokela@helsinki.fi}
Matti J\"arvinen,${}^{3,4}$\footnote{matti.jarvinen@apctp.org} and
Jere Remes${}^{5,6}$\footnote{jere.remes@iki.fi}

\vspace{2mm}
${}^1${\small \sl Department of Physics} and ${}^2${\small \sl Helsinki Institute of Physics} \\
{\small \sl P.O.Box 64} \\
{\small \sl FIN-00014 University of Helsinki, Finland}

 \vskip 0.1cm
 ${}^3${\small \sl Asia Pacific Center for Theoretical Physics and} \\
 ${}^4${\small \sl Department of Physics} \\
 {\small \sl Pohang University of Science and Technology} \\
 {\small \sl Pohang 37673, Republic of Korea} 
 
  \vskip 0.1cm
 ${}^5${\small \sl Departamento de F\'{\i}sica, Universidad de Oviedo and} \\
 ${}^6${\small \sl Instituto de Ciencias y Tecnolog\'{i}as Espaciales de Asturias (ICTEA)} \\
 {\small \sl c/ Federico Garc\'{\i}a Lorca 18, ES-33007, Oviedo, Spain}

\end{center}


\setcounter{footnote}{0}
\renewcommand\thefootnote{\mbox{\arabic{footnote}}}

\begin{abstract}
\noindent
We analyze families of hybrid equations of state of cold QCD matter, 
which combine input from gauge/gravity duality and from various ab initio methods for nuclear matter at low density, and predict that all neutron stars are fully hadronic without quark matter cores. We focus on constraints from recent measurements by the \emph{NICER} telescope on the radius and mass of the millisecond pulsar PSR J0740+6620. These results are found to be consistent with our approach: they set only mild constraints on the hybrid equations of state, and favor the most natural models which are relatively stiff at low density. Adding an upper bound on the maximal mass of neutron stars, as suggested by the analysis of the GW170817 neutron star merger event, tightens the constraints considerably. We discuss updated predictions on observables such as the transition density and latent heat of the nuclear to quark matter transition as well as the masses, radii, and tidal deformabilities of neutron stars.
\end{abstract}



\newpage

\section{Introduction}\label{sec:introduction}

In the past few years, a 
new 
window to study the behavior of matter at immense pressures and densities has been opened 
by advances in both the astrophysical observations 
of neutron stars (NS) and the modeling of their properties.  
These advances lead, in turn, 
to more stringent constraints on the equation of state (EoS) of dense matter. With better instrumentation measuring e.g. the Shapiro delay from 
pulsars locked in binary systems has become feasible, providing us with a more accurate assessment of masses of NSs. One such is the most massive NS observed to date, 
the millisecond pulsar 
PSR J0740+6620, which has a mass of $M=2.08 \pm 0.07 M_\odot$ (68\% confidence) \cite{Cromartie:2019kug,Fonseca:2021wxt}. 
Apart from improved mass measurements, 
a major breakthrough in 
measuring NS properties 
has been the LIGO/Virgo gravitational wave (GW) observation of a NS merger GW170817~\cite{TheLIGOScientific:2017qsa,Abbott:2018exr}, together with the observation of its electromagnetic counterpart~\cite{LIGOScientific:2017ync}. 
This event has opened  
the era of multimessenger astronomy and 
already provides   
constraints on the EoS by restraining the tidal deformability of compact stars \cite{Abbott:2018exr}.

Advances in theoretical modeling as well as a growing body of data from X-ray satellites have led to considerable progress  
in NS radius measurements. 
To this end, 
multiple different methods, usually employing either spectroscopic or timing information, have been developed 
in the recent years (for reviews on the subject see Refs.~\cite{Ozel:2016oaf,Miller:2016pom}). Promising results have been attained for example by analyzing the time evolution of X-ray burst cooling tail spectra from low-mass X-ray binaries (LMXBs) with the use of sophisticated atmospheric models~\cite{Nattila:2015jra,Nattila:2017wtj}. Another venue has been explored by the Neutron Star Interior Composition Explorer (\emph{NICER}) collaboration, who have used Bayesian methods in modeling the energy-dependent X-ray pulse waveform data from millisecond pulsars, with independent teams analyzing the data using different models for the X-ray emitting hot spots as well as for the instrumental response. 

In recent papers, the combined X-ray timing data~\cite{Wolff:2021oba} from \emph{NICER} and the spectroscopic data from the X-ray Multi-Mirror (XMM-Newton) on the above-mentioned pulsar PSR J0740+6620 was analyzed by two teams: by Miller et al. in~\cite{Miller:2021qha} and by Riley et al. in~\cite{Riley:2021pdl}. These analyses produced two differing but mutually compatible estimates on the equatorial radius of PSR J0740+6620, with Miller et al. finding it to be $13.7^{+2.6}_{-1.5}$ km while Riley et al. find $12.39^{+1.30}_{-0.98}$ km, both expressed at 68\% credibility. 

These various constraints on the 
masses and radii of NSs set in turn constraints to 
EoS of dense matter from compact stars.
They 
have been studied in multiple articles,
including \cite{Annala:2021gom,Pang:2021jta,Raaijmakers:2021uju,Li:2021sxb}, which take into account the most recent \emph{NICER} constraints. Since there are no first principles solutions available for QCD at the densities found within NSs, these approaches usually rely on various interpolation schemes between the known nuclear physics at low density and perturbative QCD at high density. For example, Annala et al.~\cite{Annala:2021gom} use piecewise interpolations in speed of sound squared $c_s^2$ as a function of chemical potential. 
It is, however, clear  
that even pinpointing the EoS using interpolative methods is not sufficient in determining 
definitely 
the underlying degrees of freedom in the absence of additional information \cite{Alford:2004pf}. 
Gravitational wave signals arising from NS mergers may (for example) provide evidence of first order nuclear to quark matter transition taking place during the merger~\cite{Most:2018eaw,Bauswein:2018bma,Chesler:2019osn,Ecker:2019xrw}. However, since mergers also depend mainly on the (cold) EoS, apart from possible minor contributions from bulk viscosity~\cite{Alford:2017rxf} and heating of the matter in the merger,  additional information is somewhat limited.
Therefore it is important to support the analysis of the EoS by modeling of the underlying theory, which can give additional constraints, for example, for the transport properties of the matter in the various possible phases. Moreover, using well-motivated interpolative models may unintentionally exclude EoSs with unusual features, such as abrupt changes in the speed of sound 
without any underlying reason.

In this article, we will be using gauge/gravity duality, combined with predictions from other sources, to analyze the QCD EoS at zero temperature.
We will take the new \emph{NICER} radius constraints into account and discuss their implications on the underlying EoS and what it means for the composition of NSs in the light of a state-of-the-art holographic model of QCD in the Veneziano limit.

The toolbox provided by gauge/gravity dualities is by now well-established. Methods utilizing these dualities between a strongly coupled field theory and a weakly coupled gravity theory have in recent years shown much promise in helping map the gaps between the known regions in the phase diagram. Holographic methods have been especially useful in providing insight in the high-temperature, low-density regime of quark gluon plasma produced in heavy-ion collisions~\cite{CasalderreySolana:2011us,Ramallo:2013bua,Brambilla:2014jmp}. With the improved understanding of finite-density states in holography, the models have also recently found application in NSs~\cite{Fadafa:2019euu,BitaghsirFadafan:2020otb,Kovensky:2021kzl}. Top-down models, such as the D3-D7 model~\cite{Hoyos:2016zke,Annala:2017tqz,Faedo:2017aoe} and the Witten-Sakai-Sugimoto model~\cite{Sakai:2004cn,Sakai:2005yt,Burikham:2010sw,Ghoroku:2013gja,Li:2015uea,Preis:2016fsp,Elliot-Ripley:2016uwb,BitaghsirFadafan:2018uzs,Pinkanjanarod:2020mgi,Kovensky:2021ddl}, which are rigorously based in string theory, are great in providing insights into the behavior of strongly coupled systems and the nature of the holographic duality. These results can be complemented by tailored bottom-up models, which do not have similar string theory foundation, but can be adjusted to reproduce the phenomenology of QCD at even better precision; see recent review~\cite{Jarvinen:2021jbd}. 

The V-QCD model, which we use in this work, is a phenomenologically motivated bottom-up holographic model
allowing us to  
closely mimic QCD behavior in the relevant aspects. This model, introduced in~\cite{Gursoy:2007cb,Gursoy:2007er,Jarvinen:2011qe,Alho:2013hsa}, has been recently used to study dense and cold QCD matter in~\cite{Jokela:2018ers,Ishii:2019gta,Chesler:2019osn,Ecker:2019xrw,Jokela:2020piw}. The construction of the model 
is discussed 
in detail in a recent review~\cite{Jarvinen:2021jbd}, and the basic essence of it is to tune a somewhat complex, noncritical string theory inspired five-dimensional gravity action with nontrivial potentials to first reproduce generic features of QCD and then use the available lattice data at low densities to fix the potentials. What is remarkable is that the fitting of the thermodynamics is at all possible, and that it is fairly insensitive to the exact values of the different fit parameters~\cite{Jokela:2018ers}. 
Consequently, the predictions of the model are tightly constrained. 

The approach we use to include nuclear matter in the model, which assumes a spatially homogeneous configuration~\cite{Rozali:2007rx,Ishii:2019gta}, is expected to only work at high densities. Therefore we expect that our description of nuclear matter becomes unreliable
when the baryon density drops to below around one to two times the nuclear saturation density. 
To rectify this ignorance, we use a selection of established nuclear physics models in the 
low density 
regime to supplement our holographic description, creating hybrid EoSs~\cite{Ecker:2019xrw,Jokela:2020piw} with a smooth transition between the weakly coupled and the strongly coupled descriptions of nuclear matter.

Some of our main results are the following. We notice that our model is fully consistent with all the current astrophysical constraints, 
including the latest \emph{NICER} results. We 
study 
the effect of the astrophysics constraints on the possible values of the speed of sound squared $c_s^2$, with the maximum being reached at the transition to the quark matter phase. We  
show that the hybrid EoSs constrained with the available radius, mass and tidal deformability measurements favor larger radii, with the most tightly constrained limits for the radius of 1.4 solar mass NS lying between 12.0 km and 12.8 km. 
We then study the effect of the constraints on some of the most relevant characteristic frequencies of the gravitational wave signal and note that the radius constraints considerably limit the possible values for any given mass, with lower frequencies being favored. We  
also consider the ranges for the observable parameters and notice that the latent heat in the nuclear matter to quark matter transition, happening at densities between $4.3 \lesssim n_b/n_s \lesssim 6.8$ (where $n_s \approx 0.16$ fm$^{-3}$ is the nuclear saturation density), is always in excess of 750 MeV/fm$^3$, meaning that the model does not support the existence of quark matter cores inside of stable neutron stars.

The rest of this paper is organized as follows. In Sec.~\ref{sec:methods} we will briefly overview the methods used in this paper. We will explain how the hybrid EoS is constructed. We will also explain how we take into account recent observational data. In Sec.~\ref{sec:results} we will turn to results for the EoS, for mass radius relations, and for characteristic gravitational wave frequencies of NS mergers. We will finish with a discussion in Sec.~\ref{sec:discussion}.

\section{Methods}\label{sec:methods}

In this section we will briefly explain the methods used to carry out the analysis in this paper. For more details we refer the reader to \cite{Jokela:2020piw}. The results of the analysis in the light of new astrophysics constraints will be presented in Sec.~\ref{sec:results}. 

\subsection{Equation of state}\label{subsec:eos}

As alluded to in the introduction, the equation of state for QCD matter is largely unknown and one either needs to parametrize the ignorance by, \emph{e.g.}, an interpolation scheme consistent with known perturbative limits at small and large densities, or by model calculations. In this paper we will adopt the unified weak/strong coupling framework of \cite{Jokela:2020piw} and construct hybrid EoSs describing three regions:
\begin{itemize}
 \item For nuclear matter (NM) at low densities, up to densities equal to roughly 2 times the saturation density $n_\textrm{s}=0.16$fm$^{-3}$, we choose various weakly coupled models of nuclear matter that are still plausible given astrophysical constraints, namely:
    \begin{center}
     HLPSs \cite{Hebeler:2013nza} \qquad APR \cite{Akmal:1998cf} \qquad SLy \cite{Haensel:1993zw,Douchin:2001sv} \qquad HLPSi \cite{Hebeler:2013nza} \qquad IUF \cite{Hempel:2009mc,Fattoyev:2010mx}
    \end{center}
    ordered here from soft to stiff. The HLPSs and HLPSi mean soft and intermediate variants of \cite{Hebeler:2013nza}, respectively.  
 \item For dense NM, we choose the V-QCD model with the homogeneous approach as established in \cite{Ishii:2019gta}.
 \item For quark matter (QM) we choose the V-QCD EoSs constructed in \cite{Jokela:2018ers}. The EoS for the QM is not completely unambiguous as the fitting of the holographic model to lattice QCD results at small density is not restrictive enough. The span of all realistic possibilities is, however, well represented by the choice of three distinct potentials, which will here be called:
 \begin{center}
 soft (Pot. {\bf 5b}), \qquad intermediate (Pot. {\bf 7a}), \qquad stiff (Pot. {\bf 8b}).
 \end{center}

\end{itemize}
The last two regions, as well as the transition from the NM to QM, are described by the same holographic model. 
This kind of 
hybrid EoSs were constructed in \cite{Ecker:2019xrw} in the study of numerical simulation of neutron star mergers. There is one parametric freedom. This is where we choose to match the weakly coupled models to the holographic model representing the strongly interacting sector of QCD. This parameter we denote by the transition density $n_{\text{tr}}$ \cite{Jokela:2020piw} and take it lie in the range $1.2n_\textrm{s}\ldots 2.6n_\textrm{s}$. 

We emphasize that there are several EoSs constructed in this way that are consistent with all known astrophysical constraints. To pick representatives, we have chosen to depict hybrid EoSs corresponding to APR (zero temperature nucleonic $npe\mu$ matter in $\beta$-equilibrium) matched at $n_{\text{tr}}/n_\textrm{s}=1.6$ with the holographic model. The three representatives corresponding to the different choices of potentials are notated in the figures in red as follows:
\begin{itemize}
\item V-QCD(APR) soft: dotted curve,
\item V-QCD(APR) intermediate: solid curve,
\item V-QCD(APR) stiff: dashed curve.
\end{itemize}
Notice that these EoSs are deposited in the online repository CompOSE (\underline{Comp}Star \underline{O}nline \underline{S}upernovae \underline{E}quations of State online service, \url{https://compose.obspm.fr/}).

\subsection{Constraints} \label{ssec:constraints}

Let us next discuss the astrophysical constraints that are implemented in our analysis. 

We will implement the constraint stemming from the analysis LIGO/Virgo collaborations due to the (non-)observation of the squishiness of the NS in the gravitational wave signal\cite{Abbott:2018exr} in the event GW170817. This constraint can be most easily realized as a window
\be\label{eq:ligo}
 580 \geq \Lambda(1.4M_\odot)\geq 70 \ ,
\ee
for the viable values of the tidal deformability $\Lambda$ (at 90\% confidence level), which measures how much the quadrupole moment of the NS is deformed by tidal forces \cite{Yagi:2013awa}, for a compact star of mass $1.4M_\odot$.

One constraint is due to observations of very massive NSs, such as J0348+0432 ($2.01 \pm 0.04 M_\odot$, 68\% credibility)\cite{Antoniadis:2013pzd} and J0740+6620 ($2.08 \pm 0.07 M_\odot$, 68\% credibility)\cite{Cromartie:2019kug,Fonseca:2021wxt}. In the following we will exclude any hybrid EoS if it does not support a star whose mass is at least
\be\label{eq:mass}
 M\geq 2M_\odot \ .
\ee

In this paper we also implement further constraints obtained by two independent Bayesian analyses by Riley et al.~\cite{Riley:2021pdl} and by Miller et al.~\cite{Miller:2021qha}, of the X-ray timing data on the massive millisecond pulsar PSR J0740+6620 by the \emph{NICER} experiment. Both teams combined the \emph{NICER} data along with the spectrographic data from the X-ray Multi-Mirror (XMM-Newton) to constrain the equatorial radius of PSR J0740+6620, using for example different prior assumptions and statistical sampling protocols. The Riley et al. $1\sigma$ estimate for the equatorial radius of J0470+6620 is $R_{\text J0470+6620}=12.39^{+1.30}_{-0.98}$ km, and since the upper limit turns out not to constrain the EoSs at all, this gives us a lower limit of
\be
R_{\text J0470+6620} > 11.4 \text{ km} , \label{eq:riley}
\ee
with the corresponding result for Miller et al. being $R_{\text J0470+6620} =13.7^{+2.6}_{-1.5} $ km, giving a lower limit of
\be
R_{\text J0470+6620} > 12.2 \text{ km}. \label{eq:miller}
\ee
It should be noted that we will make some simplifying assumptions when using these results;
in particular, we apply them at $M=2M_\odot$, which coincides roughly with the \emph{NICER} $1\sigma$ limit. 
We do not use the full two-dimensional marginal posterior probability obtained by the \emph{NICER} teams for the mass and equatorial radius (with these $1\sigma$ results being shown in Fig.~\ref{fig:MRbands}), but instead we use the one-dimensional, marginalized results for the radius, which are (in principle) obtained by integrating out the other variable which the marginal joint posterior density depends on. 
Using the two-dimensional probability distributions would be complicated due to the fact that some of the viable mass radius contours end due to an instability in the regime where the two-dimensional distribution is nontrivial. 
The mass of J0740+6620 was measured to be $M=2.08 \pm 0.07 M_\odot$ (68\% confidence) \cite{Cromartie:2019kug,Fonseca:2021wxt}, so projecting to the results to the median value would lead to the loss of some realistic EoSs that are also compliant with the full two-dimensional constraints, as some of them do not support masses higher than $M=2M_\odot$.
Moreover, all the relevant mass radius curves bend towards smaller radii with increasing mass in this region. 
Therefore, using a value lower -- in our case, $M=2.0 M_\odot$ -- than the median value for the mass of J0740+6620 is a more conservative approach to constraining the EoSs.

We also consider the effect of previous \emph{NICER} results for a lighter NS, PSR J0030+0451 as analyzed by Riley et al. in Ref.~\cite{Riley:2019yda} and Miller et al. in Ref.~\cite{Miller:2019cac}. For these results, the different analyzes give different posterior estimates for the mass of the star, with the median values being $1.34 M_\odot$ and $1.44 M_\odot$ respectively, as well as the equatorial radius, $12.71^{+1.14}_{-1.19}$ km and $13.02^{+1.24}_{-1.06}$ km respectively. We also indicate the J0030+0451 results in Fig.~\ref{fig:MRbands}, with Riley et al. results being noted by the blue unshaded area, and the Miller et al. results with the black unshaded area, along with the effect of implementing these constraints on the hybrid EoSs, shown on the right panel by the correspondingly colored striped bands.

In all the figures in this paper the black bands correspond to Riley et al. constraint~\eqref{eq:riley} and the blue bands correspond to the Miller et al. constraint~\eqref{eq:miller}. Older results for the lighter PSR J0030+0451 from Ref.~\cite{Riley:2019yda} (on striped black) and from Ref.~\cite{Miller:2019cac} (on striped blue) that constrain lighter NSs, with masses $\sim 1.4M_\odot$ are also included in the graphs. The older results do not play as significant a role and the most stringent constraint comes from restricting the radii of two-solar mass stars as inferred from observing PSR J0740+6620, and the EoSs that satisfy the Miller et al. constraint~\eqref{eq:miller} also satisfy all other radius constraints considered here.

One further constraint we consider in the paper comes from the analysis of the remnant of the GW170817 NS merger. There is evidence that the remnant collapsed to a black hole~\cite{Margalit:2017dij,Shibata:2017xdx,Rezzolla:2017aly,Ruiz:2017due,Shibata:2019ctb}, which gives us a hard upper limit to the possible mass of a NS, but the exact limit depends on the scenario which the collapsing body went through~\cite{Annala:2021gom}. The most conservative estimate presented in Ref.~\cite{Annala:2021gom} for the maximal mass would be $M<2.53 M_\odot$, but here we 
implement 
their stricter limit of $M<2.19M_\odot$  
in order to probe how much such limitations restrict the construction of hybrid EoSs.

\begin{figure}[!t]
	\begin{center}
\includegraphics[angle=0,width=0.95\textwidth]{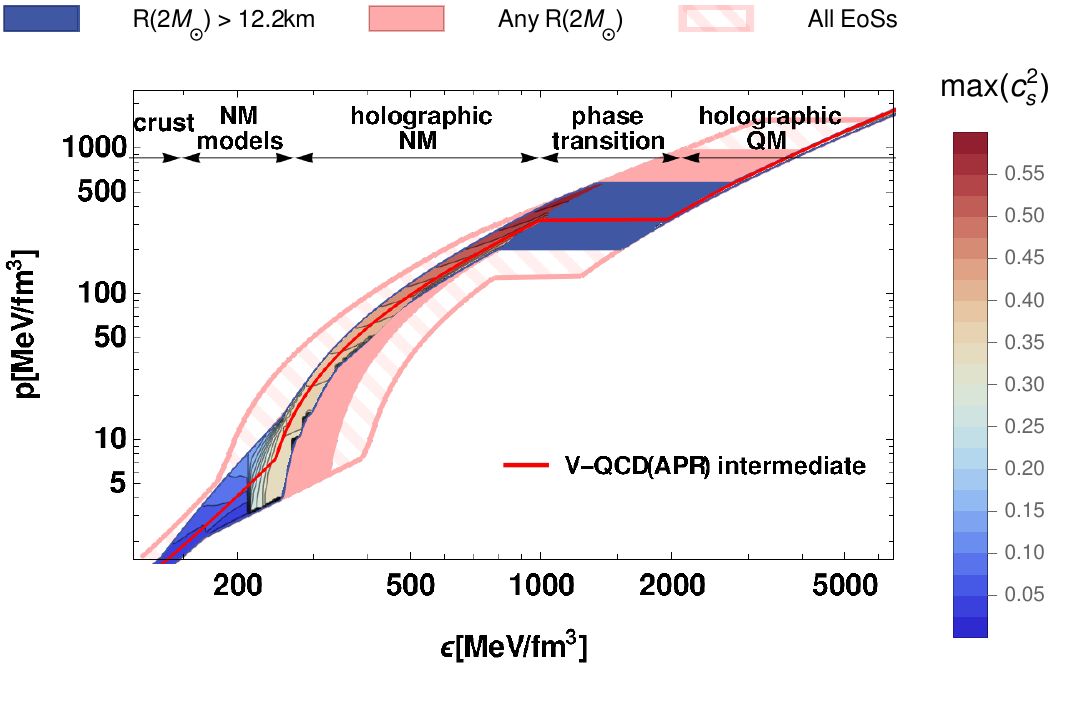}
		\end{center}\caption{The bands spanned by the hybrid EoSs in the energy density $\epsilon$ -- pressure $p$ plane, subject to different constraints, with the different regions of the EoS demarcated on the top and maximum speed of sound squared indicated by the heat map. The striped band corresponds to all of the hybrid EoSs, the light red band with the ones not subject to the \emph{NICER} radius constraints, and the blue one (with the heat map overlapped) to the most stringent \emph{NICER} radius constraint provided by Miller et al.~\cite{Miller:2021qha}. Also shown is the V-QCD (APR) intermediate hybrid EoS on red.}
	\label{fig:introfigs} 
\end{figure}

\begin{figure}[!ht]
	\begin{center}
\includegraphics[angle=0,width=0.8\textwidth]{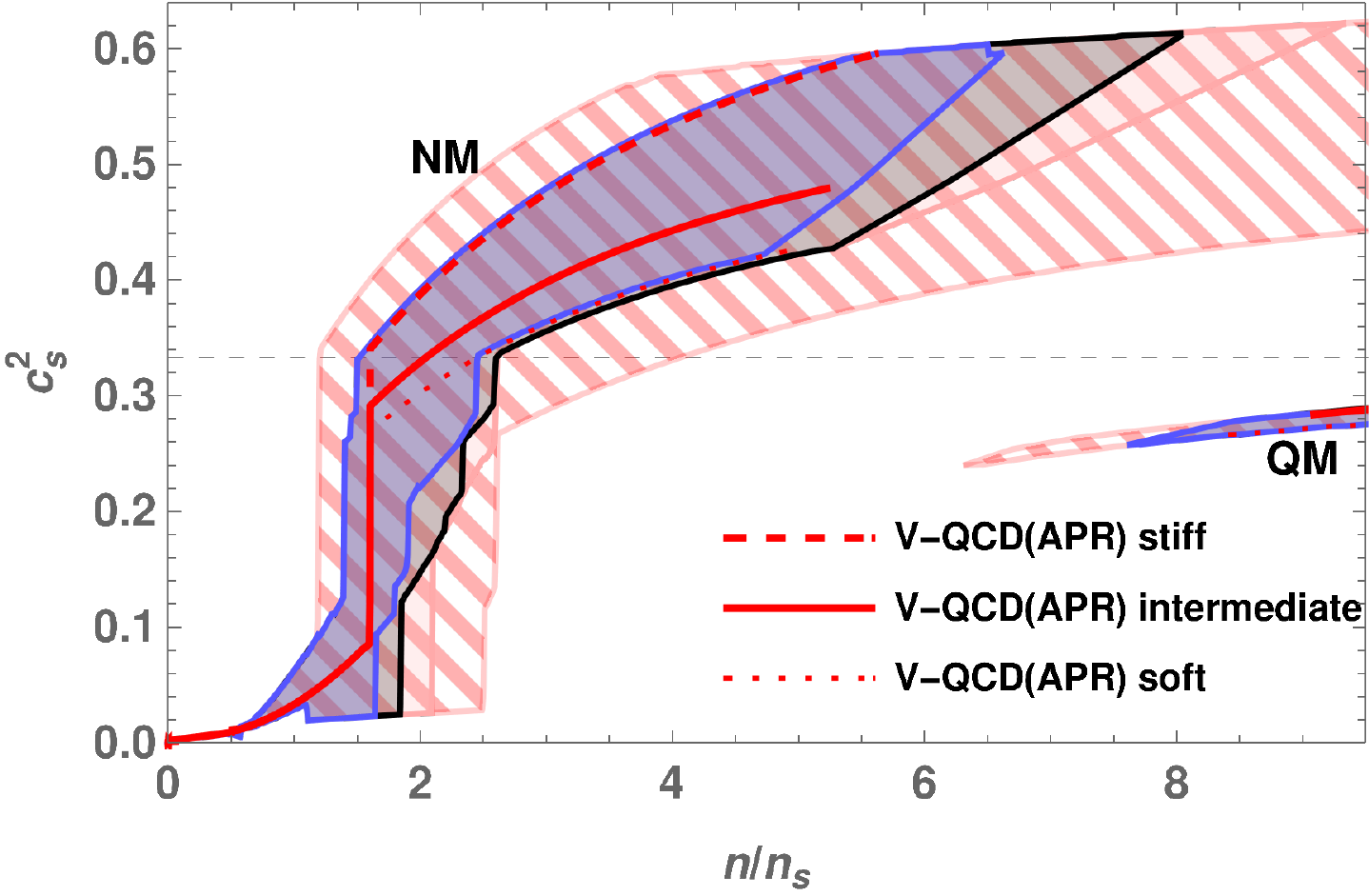}
		\end{center}\caption{The speed of sound squared $c_s^2$ as a function of number density $n$ in the units of nuclear saturation density $n_s$ as spanned by the hybrid EoSs. The different bands are as in the previous plot, with the addition of Riley et al.~\cite{Riley:2021pdl} constrained EoSs on black. The different red curves correspond to the different variants of the hybrid EoSs V-QCD (APR), with the dashed line being the variant with the stiffest V-QCD potential, solid with the intermediate and dotted with the soft, all combined with APR and with the EoSs available in CompOSE online repository. The dashed horizontal line shows the speed of sound for conformal theories.}
	\label{fig:introfigs2} 
\end{figure}

\begin{figure}[!ht]
	\begin{center}
		\includegraphics[width=\textwidth]{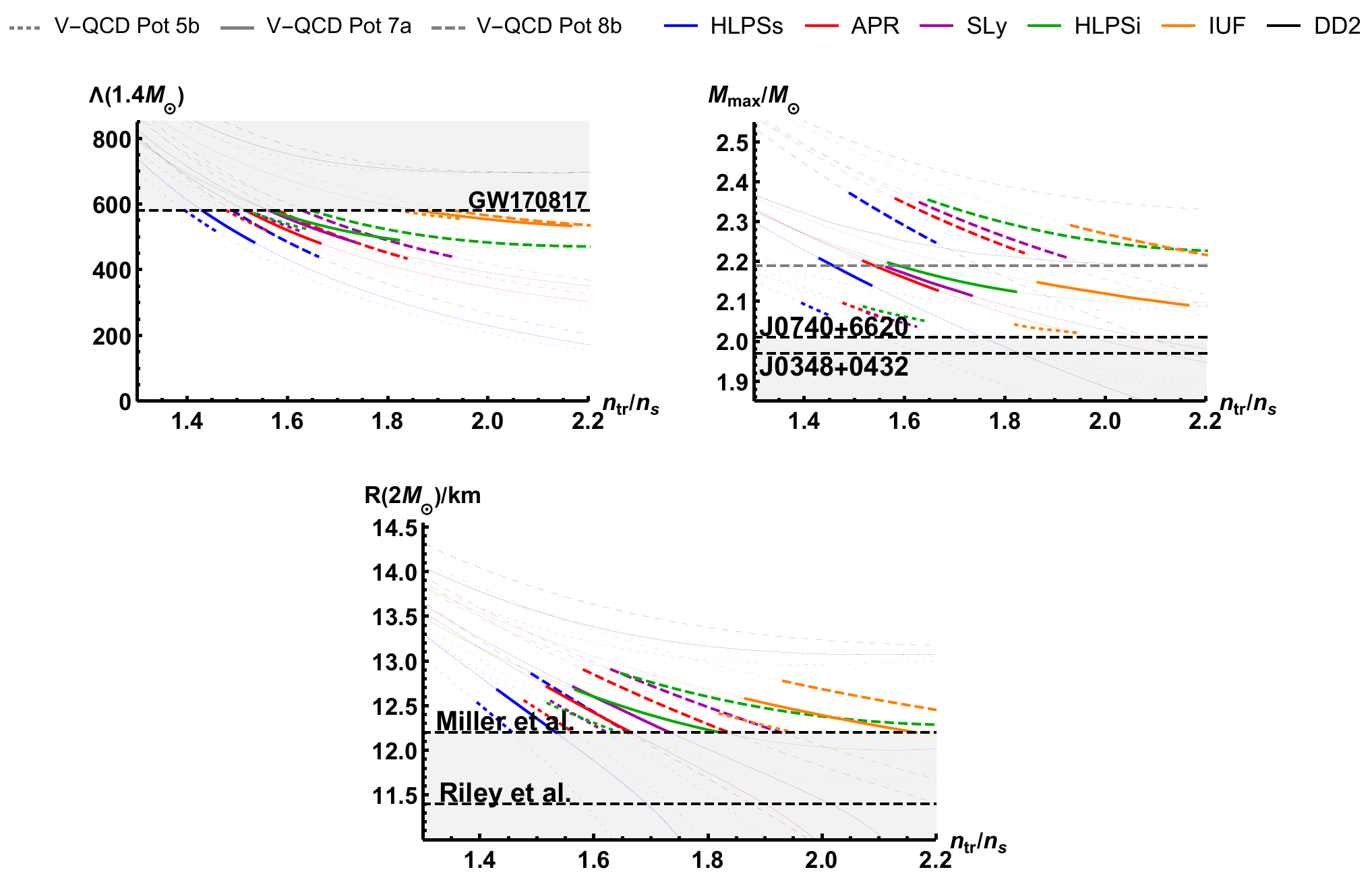}
	\end{center}
	\caption{The effect of the astrophysical constraints on $n_\textrm{tr}$ for each of the different hybrid EoSs. The areas shaded gray indicate the various constraints implemented and the thick (thin) curves correspond to the allowed (outruled) values of $n_\textrm{tr}$ for the hybrid EoSs as indicated in the legend. Top left: tidal deformability $\Lambda (1.4 M_\odot)$ as a function of $n_\textrm{tr}$ with the LIGO/Virgo bound~\eqref{eq:ligo}. Top right: maximum NS mass $M_\mathrm{max}$ in solar masses as a function of $n_\mathrm{tr}$, with the masses of J0348+0432 and J0740+6620 shown. We have also demarcated the $2.19 M_\odot$ limit for reference. Bottom: radius of a $2M_\odot$ star as a function of $n_\mathrm{tr}$ with the most recent \emph{NICER} constraints. }
	\label{fig:constr} 
\end{figure}

\section{Results}\label{sec:results} 

We then discuss the results on analyzing the hybrid EoSs that fulfill the astrophysical constraints. From any specific EoS we get to the structure of non-rotating NS by solving the Tolman-Oppenheimer-Volkov (TOV) equations. 

We start by giving an overview on the constraints for the hybrid EoSs.
In Fig.~\ref{fig:introfigs} we show the band spanned by the 
EoSs in the energy density $\epsilon$ -- pressure $p$ plane, with the different regions discussed above demarcated on the top of the plot, as well as the different colored bands showing the different constraints they satisfy and which we will discuss in Sec.~\ref{ssec:constraints}. We also show the V-QCD(APR) intermediate hybrid EoS as an example. Also shown is a heat map indicating the maximal value reached by the speed of sound squared $c_s^2$ in different regions. This can be contrasted with Fig.~\ref{fig:introfigs2}, where we show the $c_s^2$ bands, as a function of the number density $n$, spanned by the hybrid EoS -- again subject to different constraints that we will discuss later -- along with the three variations of V-QCD(APR) mentioned above. We can already note, that all the EoSs surpass the so-called conformal value of $c_s^2 = 1/3$ in the NM phase, while the maximal value stays below $c_s^2 \approx 0.6$ for all constrained EoSs. After the first order transition from NM to QM, the values are again pushed below the conformal value, which they then approach from below. This behavior in both phases is qualitatively similar to the one obtained in Refs.~\cite{Leonhardt:2019fua,Friman:2019ncm,Otto:2019zjy} using the non-perturbative methods provided by the functional renormalization group approach. Stiff EoSs in holographic nuclear matter have also been found in the Witten-Sakai-Sugimoto model by employing a homogeneous approach similar to ours~\cite{Kovensky:2021kzl}, and in a six-dimensional AdS soliton model which contains a superconducting phase~\cite{Ghoroku:2019trx,Ghoroku:2021fos}.

\begin{figure}[!ht]
	\begin{center}
		\includegraphics[width=\textwidth]{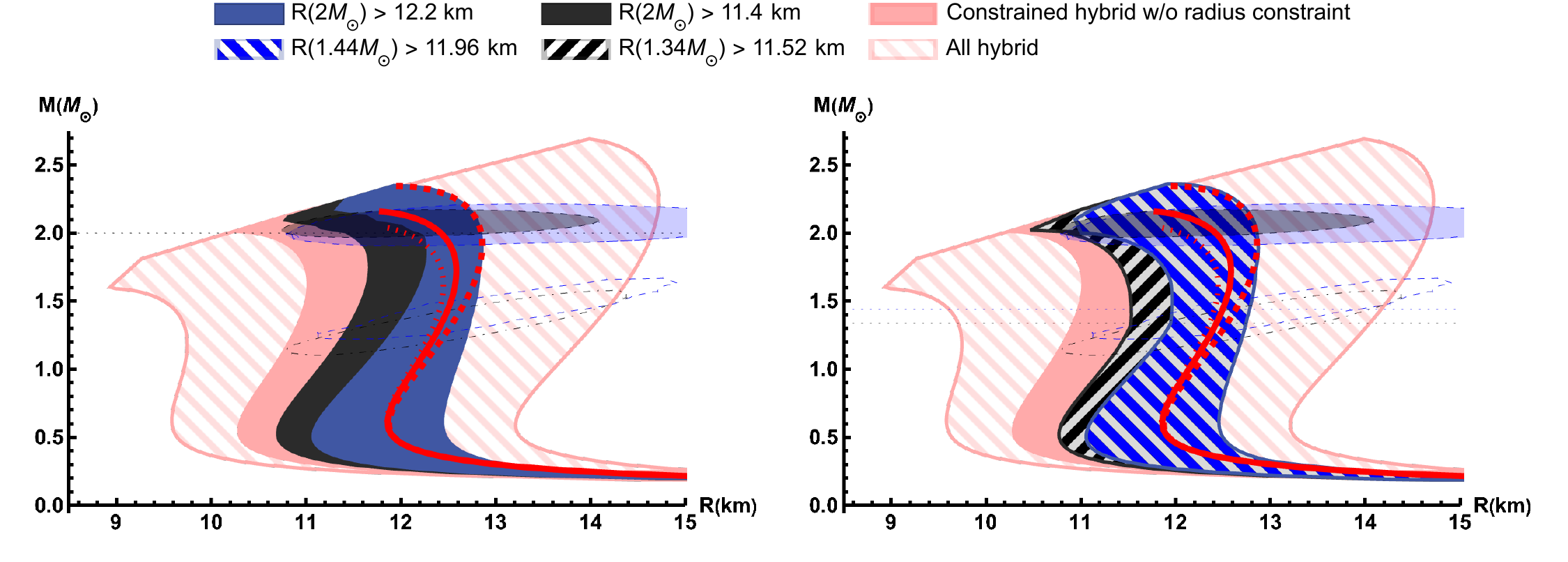}
	\end{center}\caption{Left: the Mass-Radius bands taking into account the constraints on $R(2M_\odot)$ along with constraints~\eqref{eq:ligo},~\eqref{eq:mass}. The black band corresponds to the $R(2M_\odot)>11.4$ km \emph{NICER} constraint for PSR J0740+6620 as reported by Ref.~\cite{Riley:2021pdl} (also indicated in the plot by the shaded black area) and the blue band corresponds to the $R(2M_\odot)>12.2$ km \emph{NICER} constraint as reported by Ref.~\cite{Miller:2021qha} (also indicated in the plot by the shaded blue area). The solid light red area corresponds to the non-radius constrained hybrid EoSs and the striped light red area corresponds to all non-constrained hybrid EoSs. The V-QCD(APR) curves defined in Sec.~\ref{subsec:eos}, used in Fig.~\ref{fig:introfigs2} are also presented. Also shown are the $1\sigma$ \emph{NICER} results for PSR J0030+0451 from Ref.~\cite{Riley:2019yda} (black dotdashed ellipse) and from Ref.~\cite{Miller:2019cac} (blue dashed ellipse). The horizontal dotted line marks the value $M=2M_\odot$ where we set the radius constraints for PSR J0740+6620.    
	Right: the mass-radius bands taking into account the constraints on the radius of PSR J0030+0451, with either an estimated mass of $1.34 M_\odot$ (striped black band) or $1.44 M_\odot$ (striped blue band) corresponding to the results from Ref.~\cite{Riley:2019yda} (on black) and from Ref.~\cite{Miller:2019cac} (on blue). The dotted horizontal lines show the central values for the masses from these two articles, where we set the corresponding radius constraints.
}
	\label{fig:MRbands} 
\end{figure}

We then discuss the effects of the constraints from \emph{NICER} and other sources  in detail. 
We show the effects of applying the various constraints on the parameters of the hybrid EoSs in Fig.~\ref{fig:constr}. All the astrophysical observables being constrained turn out to be descending curves as functions of the transition density $n_\textrm{tr}$, thus applying the constraints effectively brackets the allowed values of $n_\textrm{tr}$. The upper limit on $\Lambda$ brackets the values of $n_{\textrm{tr}}$ from below, while the minimum mass requirement and the minimum radius constraint for a $2 M_\odot$ star provide an upper limit on $n_{\textrm{tr}}$. As can be seen on Fig.~\ref{fig:constr}, the Miller et al. radius constraint~\eqref{eq:miller} is the more stringent of these constraints, ruling out many of the softer hybrid EoSs. 
This 
can be also seen in the left panel of Fig.~\ref{fig:MRbands}, where we have presented the $M-R$ bands spanned by the hybrid EoSs that satisfy various constraints. 
Notice, however, that the Miller et al. bound is also the most uncertain constraint due to the large spread of the two-dimensional probability distribution (with the 1$\sigma$ area shown in light blue in this plot). 
The hybrid EoSs satisfying the Miller et al. constraint of $R(2M_\odot)> 12.2$~km are marked by the blue band and the ones satisfying the Riley et al. constraint of  $R(2M_\odot)> 11.4$~km is marked by the black band. Compared to the hybrid EoSs satisfying all but the radius constraint (the light red band), the radius-constrained EoSs favor higher radii even at lower masses.

For comparison, on the right panel of Fig.~\ref{fig:MRbands} we present the $M-R$ bands spanned by the hybrid EoSs satisfying the older \emph{NICER} constraints for PSR J0030+0451. The curves satisfying the Miller et al. constraint of $R(1.44M_\odot) > 11.96$ km span the striped blue band and the ones satisfying the Riley et al. constraint of $R(1.34 M_\odot) > 11.52$ km span the striped black band.

\begin{table}[!htb]
	\begin{subtable}{0.6\textwidth}
		\centering
		\caption{$R (2 M_\odot ) > 12.2$ km}
		\begin{tabularx}{\textwidth}{c||YYYYY|}
			\hline
			\hline low density model & \footnotesize{HLPSs} & \footnotesize{APR}   & \footnotesize{SLy}   & \footnotesize{HLPSi} & \footnotesize{IUF} \\
			\hline
			\hline	min $R(1.4 M_\odot )[\text{km}]$       & 12.0 & 12.2 & 12.3 & 12.3 & 12.6 \\ 	
			max $R(1.4 M_\odot ) [\text{km}]$              & 12.6 & 12.7 & 12.7 & 12.6 & 12.8 \\ 	
			\hline	min $\Lambda (1.4 M_\odot )$           & 441  & 435  & 440  & 470  & 518  \\
			max $\Lambda (1.4 M_\odot )$                     & 580  & 580  & 580  & 580  & 580  \\
			\hline min $\Delta \epsilon$[MeV/fm$^3$]   & 762  & 780  & 751  & 760  & 747  \\
			max $\Delta \epsilon$[MeV/fm$^3$]		       & 1370 & 1430 & 1370 & 1360 & 1400 \\
			\hline        min $n_b / n_s$                            & 4.29 & 4.47 & 4.25 & 4.30 & 4.34 \\
 			max $n_b/n_s$						& 6.36 & 6.66 & 6.47 & 6.39 & 6.83 \\
		\end{tabularx}
	\end{subtable}%
	\begin{subtable}{0.4\textwidth}
		\centering
		\caption{$R (2 M_\odot )>12.2$ km \& $M < 2.19 M_\odot$}
		\begin{tabularx}{\textwidth}{|YYYYY}
			\hline
			\hline	 \footnotesize{HLPSs} & \footnotesize{APR}   & \footnotesize{SLy}   & \footnotesize{HLPSi} & \footnotesize{IUF} \\
			\hline
			\hline	 12.2  & 12.3  & 12.4  & 12.3  & 12.6  \\  	
					 12.6  & 12.7  & 12.7  & 12.6  & 12.8  \\   	
			\hline	 484   & 481   & 485   & 491   & 518   \\
					 580   & 580   & 580   & 580   & 580   \\  	
			\hline         762   & 779   & 751   & 760   & 747  \\
					 944   & 969   & 929   & 937   & 1400 \\
			\hline         4.29 & 4.47 & 4.25 & 4.30 & 4.34  \\
 					 5.28 & 5.52 & 5.25 & 5.30 & 6.83 \\
		\end{tabularx}
	\end{subtable} 
	\footnotesize
	\caption{Ranges for different parameters -- radius $R$ and tidal deformability $\Lambda$ of a 1.4 solar mass star and the latent heat $\Delta \epsilon$ and transition density $n_b$ of the NM-QM phase transition -- for NSs of fixed masses obtained from the hybrid EoSs satisfying the astrophysical constraints, with the matching density $n_\textrm{tr}/n_\textrm{s}$ ranging from 1.2 to 2.6. Even if it is not visible from this table, it is noteworthy that all the hybrid EoSs using V-QCD potential {\bf{8b}}, with the exception of IUF are ruled out when employing both the radius and the maximum mass constraints. This is due to the constraints ruling out more of the parameter space than is evident from the resulting observables alone.}\label{table:parameters}
\end{table}

Implementing the stricter \emph{NICER} constraints for the radius of a $2M_\odot$ NS affects the shape of the $M-R$ bands more than constraining the radius of the lower mass NSs: the EoSs satisfying $R(2M_\odot)> 12.2$~km also satisfy all the other radius constraints, as is also implied in subtable (a) of Table~\ref{table:parameters}. The radii for $1.4 M_\odot$ for the constrained hybrid EoSs also satisfying the $R(2M_\odot)> 12.2$~km constraint are between
\be\label{eq:R14result}
12.0 \text{ km} \leq R(1.4M_\odot)\leq 12.8 \text{ km} \ .
\ee

\begin{figure}[!ht]
	\begin{center}
		\includegraphics[width=\textwidth]{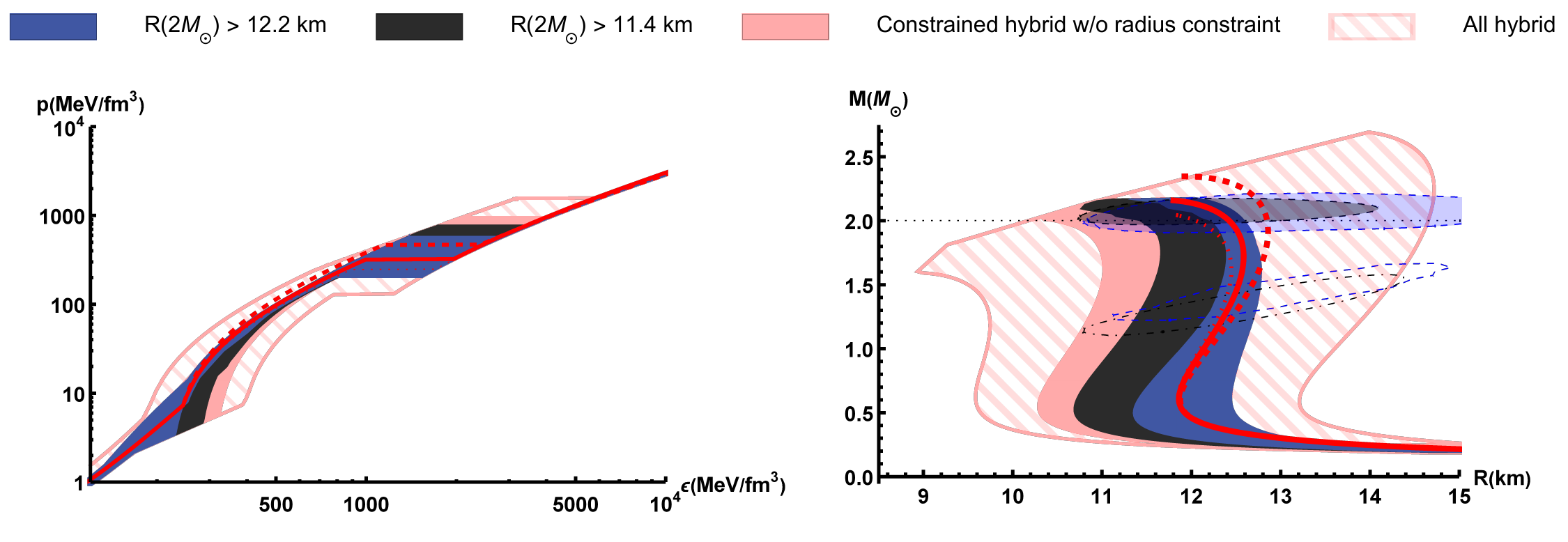}
	\end{center}\caption{Left: EoS bands spanned by the hybrid EoSs satisfying the constraints~\eqref{eq:ligo},~\eqref{eq:mass}, along with a $M<2.19 M_\odot$ constraint on the maximum mass and the different constraints on $R(2M_\odot )$. The colors of the band are consistent with the constraints in Fig.~\ref{fig:MRbands}.  
	Right: The Mass-Radius bands corresponding to the EoSs on the left, along with the various \emph{NICER} results for PSR J0740+6620 (notation as in Fig.~\protect\ref{fig:MRbands}). 
	}
	\label{fig:eosMRbands_maxM} 
\end{figure}

\begin{figure}[!ht]
	\begin{center}
		\includegraphics[width=0.8\textwidth]{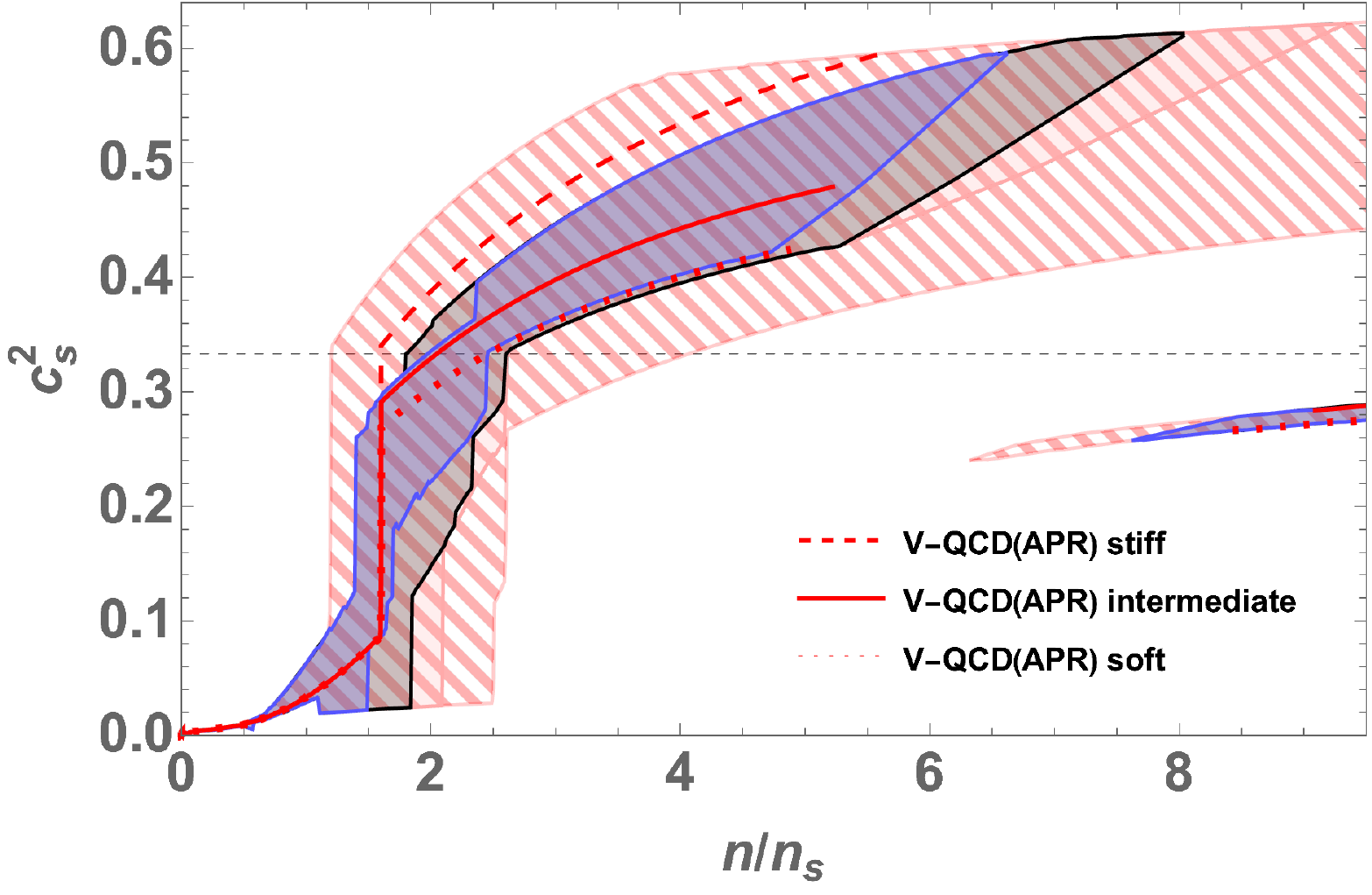}
	\end{center}
	\caption{The speed of sound squared $c_s^2$ bands spanned by the hybrid EoSs as a function of number density $n$ in the units of nuclear saturation density $n_s$. The different bands take into account the various constraints as indicated in the previous plots, as well as satisfying the maximum mass constraint of  $M<2.19 M_\odot$.}
	\label{fig:cs2_maxM} 
\end{figure}

\begin{figure}[!ht]
	\begin{center}
		\includegraphics[width=\textwidth]{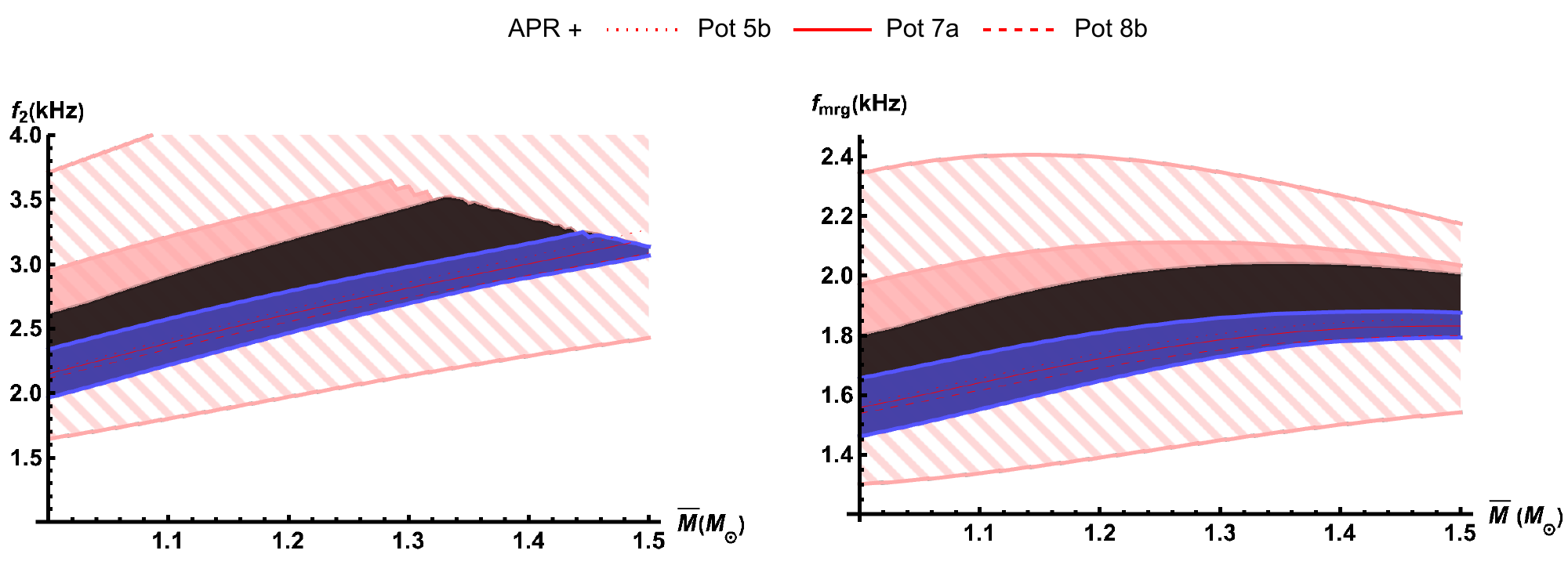}
	\end{center}
	\caption{Characteristic frequencies $f_2$ and $f_\mathrm{mrg}$ of the gravitational wave signal as functions of mass for equal  mass  binaries $M=M_A=M_B$. Also shown are example curves for APR at $n_\textrm{tr} = 1.6 n_\textrm{s}$. In the left plot we applied the prompt collapse limit $\kappa_T^2 >70$. Bands are as indicated in Fig.~\ref{fig:introfigs}}
	\label{fig:freq_M} 
\end{figure}

\begin{figure}[!ht]
	\begin{center}
		\includegraphics[width=\textwidth]{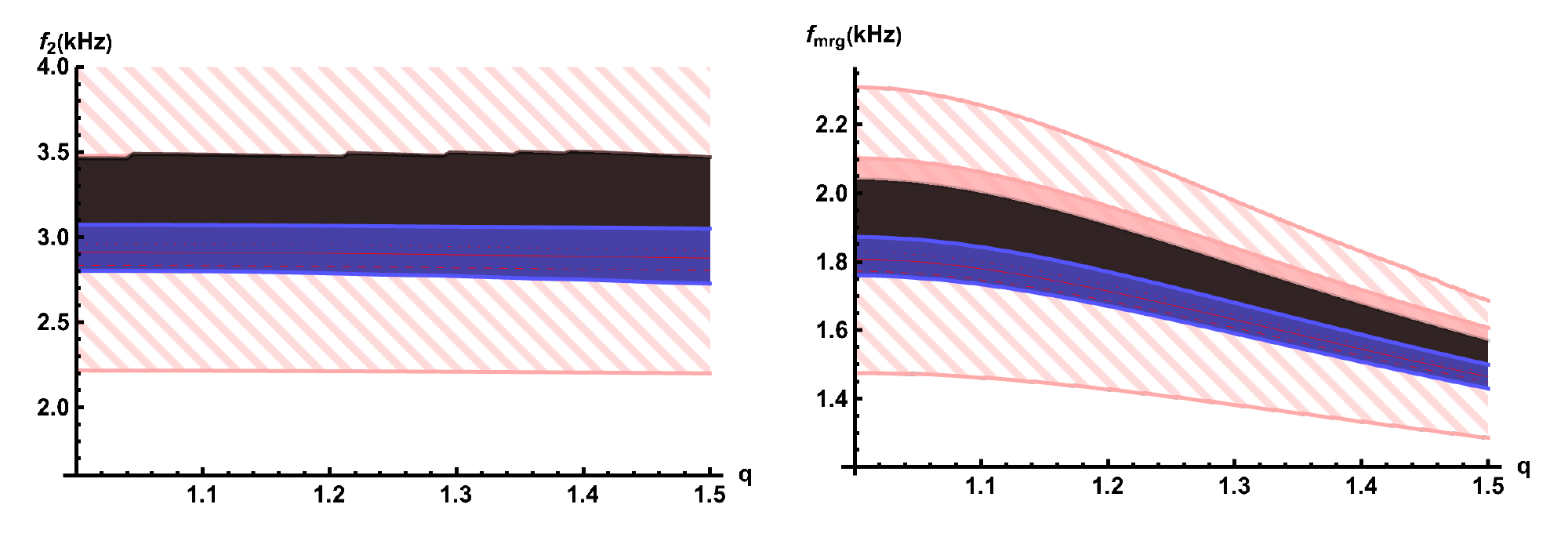}
	\end{center}
	\caption{Characteristic  frequencies  $f_2$ and $f_\mathrm{mrg}$ of  the  gravitational  wave  signal  as  functions  of $q=M_A/M_B$ for a fixed average mass of $\overline{M} = 1.35 M_\odot$.} 
	\label{fig:freq_q} 
\end{figure}

In Fig.~\ref{fig:eosMRbands_maxM} we consider the effects of further constraining the EoSs by requiring the maximum mass of the NS to conform with $M < 2.19 M_\odot$. This constraint can also be seen indicated in Fig.~\ref{fig:constr}, and we see that the additional constraint effectively suppresses all hybrid EoSs employing the stiffest V-QCD potential ({\bf 8b}), but it only slightly affects the intermediate potential ({\bf 7a}) and does not affect the soft potential ({\bf 5b}) at all. We note from Fig.~\ref{fig:eosMRbands_maxM} that the added constraint highly limits the shape of the $M-R$ band spanned by the hybrid EoSs, especially in the case where $R(2M_\odot) > 12.2$ km is also required. This is due to the fact that to satisfy the radius constraint softer NM EoSs require especially stiff cores, which leads them to more easily break the $2.19 M_\odot $ mass limit. Furthermore, already stiff EoSs have no problem satisfying the radius constraint, but tend to produce more massive stars. Therefore the combined constraints limit the stiffness of the possible EoSs quite effectively, even if this does not reflect that radically in the observables considered below. This is also reflected in Fig.~\ref{fig:cs2_maxM}, where we consider the effect of the maximum mass limit on the speed of sound squared, and note that this constraints the maximum values slightly.
 
From Table~\ref{table:parameters} we observe that the constrained hybrid EoSs favor rather large values of tidal deformability, with the values of $\Lambda (1.4 M_\odot)$ being in excess of 435 for all potentials and NM models. 
It is therefore significantly increased from the bound of 230 from~\cite{Jokela:2020piw} which assumed no input from radius measurements. Interestingly, the same number, 230, was also found recently in the Witten-Sakai-Sugimoto model in~\cite{Kovensky:2021kzl}.
Furthermore, we note that the latent heat $\Delta \epsilon$ associated with the first order transition to the quark matter phase are in excess of 740 MeV/fm$^3$ for all hybrid EoSs, which together with the fact that core pressures needed to support the existence of quark matter within NSs are in the unstable branch of solutions for the TOV equations, means that the hybrid EoSs do not support the existence of quark matter cores within quiescent NSs.

In subtable (b) of Table~\ref{table:parameters} we show the values of the same observables corresponding to a further constraint of $M < 2.19 M_\odot$. The ruling out of the stiffest EoSs is reflected in Table~\ref{table:parameters} by the raising the minimum values of $R(1.4 M_\odot)$ and $\Lambda (1.4 M_\odot)$ and lowering the maxima of $\Delta \epsilon$ for all NM models.

Along with the static properties of NSs, we also computed the characteristic frequencies of the merger and postmerger gravitational wave signal
peak frequencies $f_i$ of the power spectral density of the postmerger signal as well as the value of the instantaneous frequency at the time of the merger $f_\mathrm{mrg}$. Based on numerical merger simulations, these frequencies can be estimated to a good accuracy by certain universal relations dependent on the masses and tidal deformabilities of the participant NSs \cite{Takami:2014zpa,Takami:2014tva,Breschi:2019srl,Tsang:2019esi}. We have discussed the protocol of applying the universal relations in more detail in Ref.~\cite{Jokela:2020piw}, following mostly Ref.~\cite{Zappa:2017xba}. Here we study the effect of the constraints arising from the \emph{NICER} measurements to these results. 

In Figures \ref{fig:freq_M} and \ref{fig:freq_q} we have presented two characteristic frequencies, $f_2$ and $f_\mathrm{mrg}$, the former of which is a prominent peak in the postmerger signal linked to the rotation of the hypermassive NS created in the merger. In Fig.~\ref{fig:freq_M}, we present the frequencies as functions of mass for equal mass binaries, whereas in Fig.~\ref{fig:freq_q} we show the frequencies as functions of the mass ratio $q=M_A/M_B$ for a fixed average mass of $\overline{M}=1.35 M_\odot$. The color coding of the bands is the same as in previous figures.

The clear effect of constraining the hybrid EoSs by radius in all of the cases is to favor lower frequencies. The largest allowed 
frequencies, for an equal mass binary with $\bar{M}=1.4 M_\odot$ 
and for the EoSs satisfying the $R(2M_\odot) >12.2$ km constraint, are $f_2 \approx 3.19$ kHz and $f_\mathrm{mrg}\approx 1.89$ kHz, compared to $f_2 \approx 3.38$ kHz and $f_\mathrm{mrg} \approx 2.10$ kHz for the unconstrained EoSs.

One thing to note is that if the remnant collapses promptly into a black hole instead, the $f_2$ signal is absent, which is reflected as the sharp cut-off in the left panel of Fig.~\ref{fig:freq_q}. Due to the prompt collapse restriction, the light red and the black band overlap in the left panel of Fig.~\ref{fig:freq_q}, as can be expected by observing the left panel in Fig.~\ref{fig:freq_M}, where these bands coincide for $\overline{M}=1.35 M_\odot$.

\section{Discussion} \label{sec:discussion} 

In this article we discussed the effects of the \emph{NICER} constraints to the unified model of
dense nucleonic and quark matter model presented in \cite{Jokela:2020piw},
and demonstrated that the model is in good agreement with 
these as well as other known
astrophysical constraints. 
Moreover, 
the equations of state that we derive using holographic methods are currently 
among the few examples  
that 
apart from the observational constraints,
also satisfy theoretical constraints both at low and large densities stemming from chiral effective field theory and perturbative QCD, respectively.

As mentioned in Section~\ref{sec:methods}, our EoSs depend on the low density nuclear theory model, the choice of the matching density, and the choice of the V-QCD model. The last of these, i.e., the choice between soft, intermediate, and stiff V-QCD models is the most important source of uncertainty in the construction. The uncertainty arises from the dependence of those model parameters in the V-QCD action that are not determined by the fit to lattice data. The leftover freedom is basically a dependence on one parameter, and the soft/intermediate/stiff choices correspond to three representative values of this parameter. It would be therefore interesting if the value of this parameter could be constrained by a more careful fitting to lattice data for thermodynamics (such as higher order cumulants of the EoS) or other data such as the dependence of the EoS on magnetic field. Interestingly, a careful fit to the hadron masses carried out in~\cite{Amorim:2021gat} seems to slightly favor the intermediate choice over the soft and stiff choices. 

In the coming years, even more data on the properties of compact stars is expected and it is a realistic scenario 
that we will end up in a situation where most of the mass-radius relationship will be determined to a high degree. It is therefore tempting to ask if we then could also understand the composition of the neutron stars adequately. A precise knowledge of the mass--radius relation might reveal a kink (or potentially many kinks) that should be interpreted as a phase transition.
However,  
experimental efforts alone even in the situation with arbitrarily accurate data, which would allow a precise inversion to obtain the ``microscopic'' equation of state \cite{Alford:2004pf}, cannot definitely determine the nature of the underlying phases. One ends up knowing little beyond equilibrium. Theoretical input is therefore required. Using gauge/gravity duality to derive predictions out-of-equilibrium is a tempting alternative. In this article, we focused only on the EoS and observables depending on it, but predictions for observables beyond the EoS, such as basic transport properties, have already been analyzed in~\cite{Hoyos:2020hmq,Hoyos:2021njg}.

Lastly, we remind that our neutron stars are fully hadronic; all stars with quark matter cores are unstable due to the strong first order nuclear to quark matter transition. That is, we present a family of EoSs, which meets all constraints from neutron star measurements with fully hadronic neutron stars with speeds of sounds that mildly exceed the conformal value of $c_s^2=1/3$. Therefore our result seems to contradict the conclusions of~\cite{Annala:2019puf}, i.e., that quark matter cores are necessary to explain current observations, unless the speed of sound reaches very high values $c_s^2>0.7$. The reason for this contradiction turns out~\cite{Jokela:2020piw} to be the classification of nuclear matter: our setup predicts exceptionally low values of the adiabatic index $\gamma= d\log p/d\log \eps$ for dense nuclear matter which are below the limit value of 1.75 used in~\cite{Annala:2019puf} so that our dense nuclear matter would be classified as quark matter in their setup. Interestingly, such low values of $\gamma$ were also found recently for nuclear matter in the Witten-Sakai-Sugimoto model~\cite{Kovensky:2021kzl}\footnote{Interestingly, such low values of the adiabatic index for dense nuclear matter have also been found in an approach based on Skyrmions~\cite{Paeng:2017qvp,Ma:2020hno}.}. Apart from this small detail, we remark that our band in Fig.~\ref{fig:introfigs} is in excellent agreement with those bands of~\cite{Annala:2019puf,Annala:2021gom} that assume a strict bound for the maximum value for the speed of sound.


\paragraph{Acknowledgments}
\addcontentsline{toc}{section}{Acknowledgments}
We would like to thank Aleksi Vuorinen for discussions and comments on the draft version of this manuscript. N.~J. has been supported in part by the Academy of Finland grant no. 1322307. The work of M.~J. was supported in part by an appointment to the JRG Program at the APCTP through the Science and Technology Promotion Fund and Lottery Fund of the Korean Government. M.~J. was also supported by the Korean Local Governments -- Gyeongsangbuk-do Province and Pohang City, and by the National Research Foundation of Korea (NRF) funded by the
Korean government (MSIT) (grant number 2021R1A2C1010834). J.~R. was supported by the Finnish Cultural Foundation. We acknowledge the support from CNRS through the PICS program as well as from the Jenny and Antti Wihuri Foundation.
\noindent

\bibliographystyle{JHEP}
\bibliography{refs} 
\end{document}